\documentclass[a4paper,10pt]{article}

\usepackage{mathptmx}       
\usepackage{helvet}         
\usepackage{courier}        
\usepackage{type1cm}        

\usepackage{authblk}
\usepackage{framed}
\usepackage{makeidx}         
\usepackage{graphicx}        
\usepackage{multicol}        
\usepackage[bottom]{footmisc}

\makeindex             

\usepackage{color}

\date{}
\begin{document}
\newcommand{\avg}[1]{\langle{#1}\rangle}
\newcommand{\Avg}[1]{\left\langle{#1}\right\rangle}
\def\be{\begin{equation}}
\def\ee{\end{equation}}
\def\bc{\begin{center}}
\def\ec{\end{center}}
\def\bea{\begin{eqnarray}}
\def\eea{\end{eqnarray}}
\title{Models, Entropy and Information of Temporal Social Networks \footnote{This work has been published as a book chapter in Temporal Networks, P. Holme, and J. Saram\"aki (Eds.) Understanding Complex Systems Series, Springer (2013)}}
\author[a]{Kun Zhao}
\author[b]{M\'arton Karsai}
\author[c]{Ginestra Bianconi}
\affil[a]{Physics Department, Northeastern University, Boston 02115 MA, USA}
\affil[b]{BECS, School of Science, Aalto University, Aalto, Finland}
\affil[c]{School of Mathematical Sciences, Queen Mary University of London, London E1, 4NS, UK}
%
%
\maketitle

\begin{abstract}Temporal social networks  are characterized by  {heterogeneous} duration of contacts, which can  either follow a power-law distribution, such as in face-to-face interactions, or  a Weibull distribution, such as in mobile-phone communication. Here we model the dynamics of face-to-face  interaction and mobile phone communication by a reinforcement dynamics, which explains  the data observed in these different types of social interactions. We quantify the  information encoded in the  dynamics of these networks  by the  entropy of temporal  networks. Finally, we  show evidence that human dynamics is able to modulate the information present in social network dynamics when it follows circadian rhythms and when it is interfacing with a new technology such as the mobile-phone communication technology.
\end{abstract}

\section{Introduction}

The theory of complex networks   \cite{Dorogovtsev:2003, Newman:2003, Boccaletti:2006, Caldarelli:2007, Doro_review,Barrat:2008} has flourished thanks to the availability of new datasets on large complex systems ,such as the Internet or the interaction networks inside the cell.
In the last ten years attention has been focusing  mainly on static or growing  complex networks, with little emphasis on the rewiring of the links. The topology of these networks and their modular structure   \cite{Fortunato, Palla:2007,Lehmann, Bianconi:2009} are able to affect the dynamics taking place on them   \cite{Doro_review, Barrat:2008,Ising,Ising_spatial}.
 Only recently   temporal networks   \cite{Holme:2005,Latora:2009,Havlin:2009,Cattuto:2010, Isella:2011,Holme:2012}, dominated by the dynamics of rewirings, are starting to attract the attention of quantitative scientists working on complexity.
One of the most beautiful examples of temporal networks are social interaction networks.
Indeed, social networks  \cite{Granovetter:1973, Wasserman:1994} are intrinsically dynamical  and social interactions are continuously formed and  dissolved.
Recently we are gaining new insights into the structure and dynamics of these temporal social networks, thanks to the availability of a new generation of datasets recording the social interactions of the fast time scale. In fact, on one side we have data on face-to-face interactions coming from mobile user devices technology    \cite{Eagle:2006,Hui:2005}, or Radio-Frequency-Identification-Devices   \cite{Cattuto:2010,Isella:2011}, on the other side, we have extensive datasets on mobile-phone calls   \cite{Onnela:2007} and agent mobility   \cite{Brockmann:2006, Gonzalez:2008}.

This new generation of data has changed drastically the way we look at social networks. In fact, the adaptability of social networks is well known and several models have been suggested for the dynamical formation of social ties and the emergence of connected societies   \cite{Bornholdt:2002,Marsili:2004,Holme:2006,MaxiSanMiguel:2008}. Nevertheless, the strength and  nature of a social tie remained difficult to quantify for several years despite the careful sociological description by Granovetter \cite{Granovetter:1973}. Only recently, with the availability of data on social interactions and their dynamics on the fast time scale, it has become possible to assign to each acquaintance the strength or weight of the social interaction quantified as the total amount of time spent together by two agents in a given time window   \cite{Cattuto:2010}.

The recent data revolution in social sciences is not restricted to data on social interaction but concerns all human activities   \cite{Barabasi:2005,Vazquez:2006,Rybski:2009, Amaral:2009}, from financial transaction to mobility. From these new data on human dynamics  evidence is emerging that human activity is  bursty  and is  not described by Poisson processes     \cite{Barabasi:2005, Vazquez:2006}. Indeed, a universal pattern   of bursty activities was observed in human dynamics  such as broker activity, library loans or email correspondence. Social interactions are not an exception, and there is evidence that face-to-face interactions have a distribution of duration well approximated by a power-law   \cite{Cattuto:2010, Scherrer:2008,Stehle:2010, Zhao:2011, Karsai:2012} while they remain modulated by circadian rhythms   \cite{Karsai:2011b}.
The bursty activity of social networks has a {significant} impact on dynamical processes defined on networks   \cite{Vazquez:2007, Karsai:2011a}{.}
Here we compare these {observations} with data coming from a large dataset of mobile-phone communication   \cite{PlosOne, Frontiers} {and show} that human social interactions, when mediated by a technology, such as the mobile-phone communication, demonstrate the adaptability of human behavior. Indeed, the distribution of duration of calls does not follow any more  a power-law distribution but has a characteristic scale determined by the weights of the links, and  is described by a Weibull distribution. {At} the same time, however,  this distribution remains bursty and strongly deviates from a Poisson distribution.
We will show that both the power-law distribution of durations of social interactions and the Weibull distribution of durations and social interactions observed respectively  in face-to-face interaction datasets and in mobile-phone communication activity can be explained phenomenologically by a model with a reinforcement dynamics   \cite{Stehle:2010, Zhao:2011,PlosOne, Frontiers} responsible for the deviation from a pure Poisson process.
In this model, the longer two agents interact, the smaller is the probability that they split apart, and the longer an agent is non interacting, the less likely it is that he/she will start a new social interaction.
We observe here that this framework is also necessary to explain the group formation in simple animals   \cite{Bisson}. This suggests that the reinforcement dynamics of social interactions, much like the Hebbian dynamics, might have a neurobiological foundation. Furthermore, this is   supported  by the results on the bursty mobility of rodents   \cite{Chialvo} and on the recurrence patterns of words encountered in online conversations   \cite{Motter}.
We have therefore found ways to quantify the adaptability of human behavior to different technologies.
We observe here that this change of behavior corresponds to the very fast time dynamics of social interactions and it is not related to macroscopic change of personality consistently with the results of \cite{Lambiotte2} on online social networks.

Moreover, temporal social networks encode information   \cite{Cover:2006} in their structure and  dynamics.
This information is necessary for  efficiently navigating  \cite{Kleinberg,WS} the network, and to build collaboration networks  \cite{Newman:2001} that are  able to enhance the performance of a society.
Recently, several authors have focused on measure{s} of entropy and information for networks.
The entropy of network ensembles is able to quantify the information encoded in  a structural feature   of networks such as the degree sequence, the community structure, and the physical embedding of the network in a geometric space   \cite{Bianconi:2008,Anand:2009, Bianconi:2009}. The entropy rate of a dynamical process on the networks, such a biased random walk,  are also able to characterize the interplay between structure of the networks and the dynamics occurring on them \cite{Latora_biased}. Finally, the mutual information for the  data of email correspondence was shown to be fruitful in characterizing the community structure of the networks   \cite{Eckmann:2004} and the entropy  of human mobility was able to set the limit of predictability of human movements   \cite{Song:2010}.

Here we will characterize the entropy of temporal social networks as a proxy to characterize the predictability of the dynamical nature of social interaction networks.
This entropy will quantify how many typical configurations of social interactions we expect at any given time, given the history of the network dynamical process.
We will evaluate this entropy on a typical day of mobile-phone communication directly from data showing modulation of the dynamical entropy during the circadian rhythm. Moreover we will show that
when the distribution of duration of contacts changes from a power-law distribution to a Weibull distribution the level of information and the value of the dynamical entropy significantly change indicating that human adaptability to new technology is a further way to modulate the information content of dynamical social networks.

\section{Temporal social networks and the distribution of duration of contacts}
\label{sec:1}

\begin{figure}[h!]
\begin{center}
\includegraphics[width=3.27in ,height=80mm]{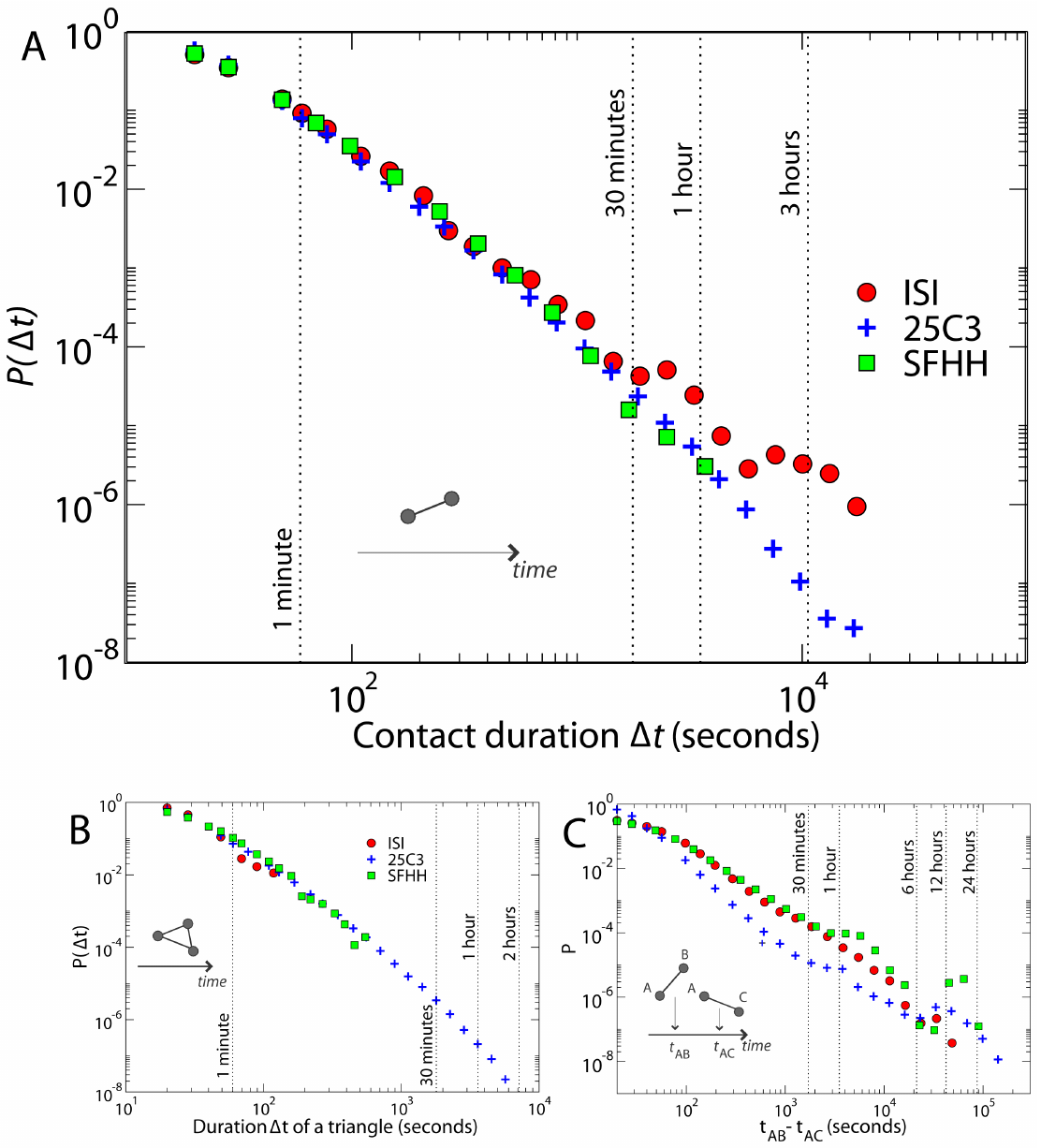}
\end{center}
\caption{ \small{Probability  distribution of human social interaction. Figure from   \cite{Cattuto:2010}
A) Probability distribution of duration of contacts between any two given persons. Strikingly, the distributions show a similar long-tail behavior independently of the setting or context where the experiment took place or the detection range considered. The data correspond to respectively 8700, 17000 and 600000 contact events registered at the ISI, SFHH and 25C3 deployments. B) Probability distribution of the duration of a triangle. The number of triangles registered are 89, 1700 and 600000 for the ISI, SFHH and 25C3 deployments. C){.} Probability distribution of the time intervals between the beginning of consecutive contacts AB and AC. Some distributions show spikes (i.e., characteristic timescales) in addition to the broad tail; for instance, the 1 h spike in the 25C3 data may be related to a time structure to fix appointments for discussions.}
}
\label{barrat}
\end{figure}

Human social dynamics is bursty, and the distribution of inter-event times follows a universal trend showing power-law tails. This is true for e-mail correspondence events, library loans,and  broker activity. 
Social interactions are not an exception to this rule, and the distribution of inter-event time between face-to-face social interactions has power-law tails   \cite{Barabasi:2005,Vazquez:2006}. Interestingly enough,  social interactions have an additional ingredient with  respect to other human activities. While sending an email can be considered an instantaneous event characterized by the instant in which the email is sent, social interactions have an intrinsic duration which is a proxy of the strength of a social tie.
In fact, social interactions are the microscopic structure of social ties and a tie can be quantified as the total time two agents interact in a given time-window.
New data on the fast time scale of social interactions have been now gathered with different methods which range from Bluetooth sensors   \cite{Eagle:2006}, to the new generation of Radio-Frequency-Identification-Devices   \cite{Cattuto:2010,Isella:2011}.
In all these data there is evidence that face-to-face interactions have a duration that follows a distribution with a power-law tail.
Moreover, there is also evidence that the inter-contact times have a distribution with fat tails.
In this chapter we report a figure of Ref.   \cite{Cattuto:2010} (Fig. \ref{barrat} of this chapter ) in which the duration of contact in Radio-Frequency-Device experiments conducted by Sociopatterns experiments  is clearly fat tailed and well approximated by a power-law (straight line on the log-log plot).
In this figure the authors of Ref.   \cite{Cattuto:2010} report the distribution of the duration of binary interactions and the distribution of duration of a the triangle of interacting agents. Moreover they report data  for the distribution of inter-event time.

How do these distributions change when human agents are interfaced with a new technology?  This is a major question that arise{s} if we want to characterize the universality of these distributions.
In this book chapter we report an  analysis of mobile-phone data and we show evidence of human adaptability to a new technology.

\begin{figure}[h!]
\begin{center}
\includegraphics[width=70mm, height=65mm]{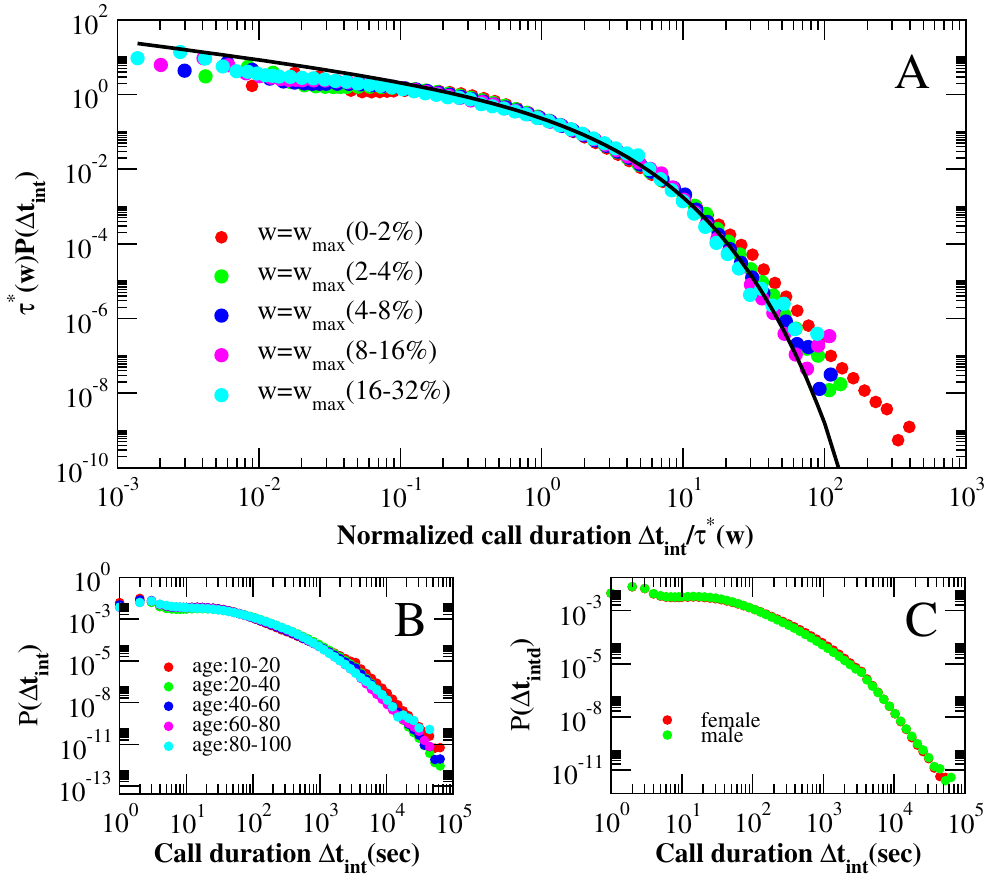}
\end{center}
\caption{ \small{(A) Distribution of duration of phone-calls between two users with weight $w$. The data depend on the typical scale $\tau^{\star}(w)$ of duration of the phone-call.
(B) Distribution of duration of phone-calls for people of different age. (C) Distribution of duration of phone-calls for users of different gender. The distributions shown in the panel (B) and (C) do not significantly depend on the attributes of the nodes. Figure from   \cite{PlosOne}.}}
\label{interaction}
\end{figure}

\begin{figure}[h!]
\begin{center}
\includegraphics[width=70mm, height=60mm]{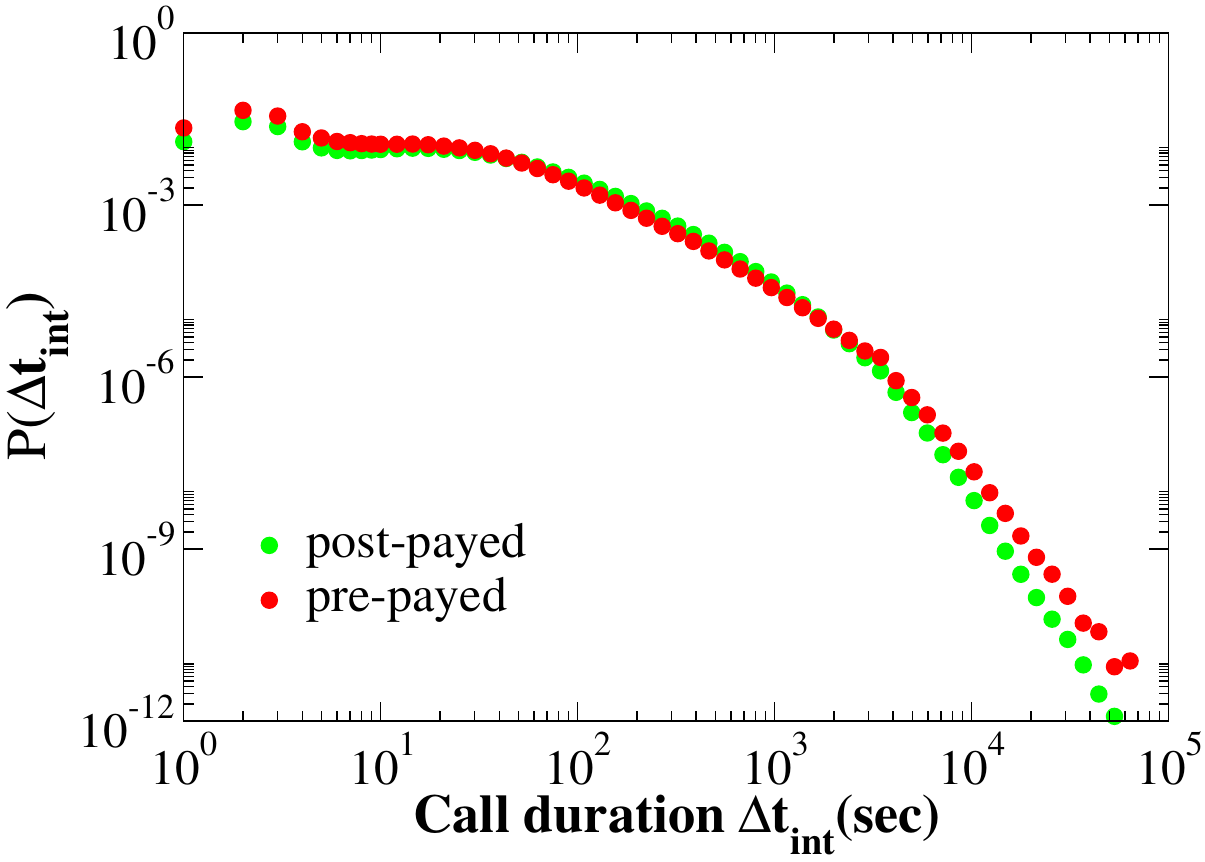}
\end{center}
\caption{  \small{Distribution of duration of phone-calls for people with different types of contract.  No significant change is observed that modifies the functional form of the distribution. Figure from   \cite{PlosOne}.}}
\label{pay}
\end{figure}

\begin{figure}[h!]
\begin{center}
\includegraphics[width=110mm, height=45mm]{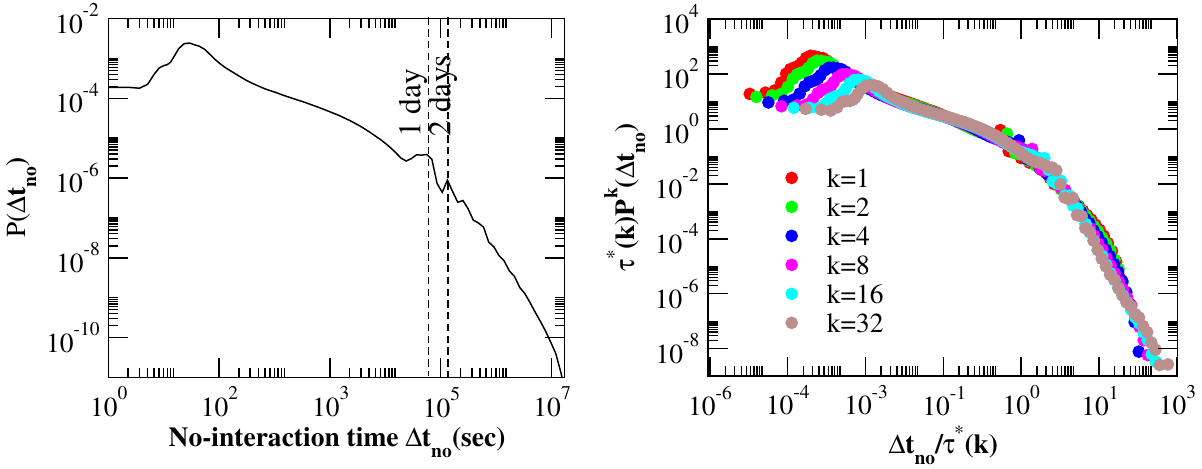}
\end{center}
\caption{ \small{Distribution of non-interaction times in the phone-call data. The distribution strongly depends on circadian rhythms. The distribution of rescaled time depends strongly on the connectivity of each node. Nodes with higher connectivity $k$ are typically non-interacting for a shorter  typical time scale $\tau^{\star}(k)$. Figure from   \cite{PlosOne}.}}
\label{non-interaction}
\end{figure}

We have analysed the call sequence of subscribers of a major European mobile service provider. In the dataset the users were anonymized and impossible to track. We considered calls between users who  called each other mutually at least once during the examined  period of $6$ months in order to examine calls only reflecting trusted social interactions. The resulted event list consists of $633,986,311$ calls between $6,243,322$ users. We have performed measurements {for} the distribution of call duration{s} and non-interaction time{s} {of} all the users for the entire 6 months time period. The distribution of phone call durations strongly deviates from a fat-tail distribution.
In Fig. \ref{interaction} we report these distributions and show that {they}  depend on the strength $w$ of the interactions (total duration of contacts in the  observed period) but do not depend on the age, gender or type of contract in a significant way.
The distribution $P^w(\Delta t_{in})$ of duration of contacts within agents with strength $w$ is well fitted by a Weibull distribution
\begin{equation}
\tau^*(w) P^w(\Delta t_{in})=W_{\beta}\left(x=\frac{\Delta t}{\tau^{\star}(w)}\right)= \frac{1}{x^{\beta}} e^{-\frac{1}{1-\beta}x^{1-\beta}}.
\end{equation}
with $\beta=0.47..$.
The typical times of interactions between users   $\tau^*(w)$ depend on the  weight $w$ of the social tie. In particular the values used for the data collapse of Figure 3 are listed in Table \ref{tauw}.
These values are broadly distributed, and there is evidence that such heterogeneity might depend on the geographical distance between the users \cite{Lambiotte}.
The Weibull distribution strongly deviates from a power-law distribution to the extent that it is characterized by a typical time scale $\tau(w)$, while power-law distribution does not have an associated characteristic scale.
 The origin  of this  significant change in the behavior of humans interactions could be due to the consideration of the cost of the interactions although we are not in the position to draw these conclusions (See Fig. \ref{pay} in which we compare distribution of duration of calls for people with different type of contract)  or might depend on the different nature of the communication.
The duration of a phone call is quite short and is not affected significantly by the circadian rhythms of the population.
On the contrary, the duration of no-interaction periods is strongly affected by the periodic daily of weekly rhythms.
In Fig. \ref{non-interaction} we report the distribution  of duration of no-interaction periods in the day periods between 7AM and 2AM next day. { The typical times $\tau^*(k)$ used in Figure 5  are listed in Table \ref{tauk}.} 
The distribution of non-interacting times is difficult to fit due to the noise derived by the dependence on circadian rhythms. In any case the non-interacting time distribution  if it is clearly fat tail.

\begin{table}
\caption{Typical times $\tau^{\star}(w)$ used in the data collapse of Fig. \ref{interaction}.}
\label{tauw}       
%
%
\begin{tabular}{p{5.5cm}p{5.5cm}}
\hline\noalign{\smallskip}
Weight of the link & Typical time $\tau^{\star}(w)$ in seconds (s)\\
(0-2\%) \ \ \ $w_{max}$ &111.6 \\
(2-4\%) \ \ \ $w_{max}$ & 237.8 \\
(4-8\%) \ \ \ $w_{max}$ & 334.4 \\
(8-16\%)\ \  $w_{max}$  & 492.0 \\
(16-32\%) $w_{max}$ & 718.8 \\
\end{tabular}
\end{table}

\begin{table}
\caption{Typical times $\tau^{\star}(k)$ used in the data collapse of Fig. \ref{non-interaction}.}
\label{tauk}       
%
%
\begin{tabular}{p{5.5cm}p{5.5cm}}
\hline\noalign{\smallskip}
Connectivity & Typical time $\tau^{\star}(k)$ in seconds (s) \\
k=1 &158,594 \\
k=2 &118,047 \\
k=4 & 69,741 \\
k=8 & 39,082 \\
k=16 & 22,824 \\
k=32 & 13,451 \\
\end{tabular}
\end{table}

\section{Model of social interaction}

It has been recognized that human dynamics is not Poissonian. Several models have been proposed for   explaining a fundamental case study of this dynamics, the data on email correspondence.
The two possible explanations of bursty email correspondence are described in the following.
\begin{itemize}
\item
A queueing model of tasks with different priorities has been suggested to explain bursty interevent time. This model implies rational decision making and correlated activity patterns \cite{Barabasi:2005,Vazquez:2006}. This model gives rise to power-law distribution of inter event times.
 \item
A convolution of Poisson processes due to different activities during the circadian rhythms and weekly cycles have been suggested to explain bursty inter event time. These different and multiple  Poisson processes are  introducing  a set of distinct characteristic time scales on human dynamics giving rise to fat tails of interevent times \cite{Malmgren}.
 \end{itemize}

In the previous section we have showed evidence that the duration of social interactions is  generally non Poissonian.
Indeed, both the power-law distribution observed for duration of face-to-face interactions and the Weibull distribution observed for duration of mobile-phone communication
strongly deviate from an exponential.
The same can be stated for the distribution of duration of non-interaction times, which strongly deviates from an exponential distribution both for face-to-face interactions and for mobile-phone communication.
In order to explain the data on duration of contacts we cannot use any of the models proposed for bursty interevent time in email correspondence.
In fact, on one side it is unlikely that the decision to continue a conversation depends on  rational decision making. Moreover the queueing model \cite{Barabasi:2005, Vazquez:2006} cannot explain the observed stretched exponential distribution of duration of calls. On the other side, the duration of contacts it is not effected by circadian rhythms and weekly cycles which are responsible for bursty behavior in the model  \cite{Malmgren}.
This implies that a new theoretical framework is needed to explain social interaction data.
Therefore, in order to model the temporal social networks we have to abandon the generally considered assumption that social interactions are generated by a Poisson process.
In this assumption the probability for two agents to start an interaction or to end an interaction is constant in time and not affected by the duration of the social interaction.

{ Instead, to build a model for human social interactions we have to consider a reinforcement dynamics, in which the probability to start an interaction depends on how long an individual has been non-interacting, and the probability to end an interaction depends on the duration of the interaction itself.
Generally, to model the human social interactions, we can consider an agent-based system consisting of $N$ agents that can dynamically interact with each other and give rise to interacting agent groups. In the following subsections we give more details on the dynamics of the models.  We denote by the state  $n$ of the agent, the number of agents in his/her group (including itself). In particular we notice here that a state $n=1$ for an agent, denotes the fact that the agent is non-interacting. A reinforcement dynamics for such system is defined in the following frame. }


\begin{framed}
\hspace{-.25in} {\bf Reinforcement dynamics in temporal social networks}\\
The longer an agent is interacting in a group the smaller is the probability that he/she will  leave the group.\\
 The longer an agent is non-interacting the smaller is the probability that he/she will form or join a new group.\\
The probability that an agent $i$ change his/her state (value of $n$) is given by 
\begin{equation}
f_n(t,t_i)=\frac{h(t)}{(\tau+1)^{\beta}}
\label{f}
\end{equation}
where $\tau:=(t-t_i)/N$, $N$ is the total number of agents in the model and $t_i$ is the last time the agent $i$ has changed his/her state, and $\beta$ is a parameter of the model.
The reinforcement mechanism is satisfied by any function $f_n(t,t_i)$ that is decreasing with $\tau$ but social-interaction  data currently available are reproduced only for this particular choice $f_n(t,t_i)$.
\end{framed}

The function $h(t)$ only depends on the actual time in which the decision is made. This function is able to modulate the activity during the day and throughout the weekly rhythms. For the modelling of the interaction data we will first assume that the function $h(t)$ is a constant in time. 
Moreover in  the following subsections we will show that in order to obtain power-law distribution of duration of contacts and non-interaction times (as it is observed in face-to-face interaction data) we have to take $\beta=1$ while in order to obtain Weibull distribution of duration of contacts we have to take $\beta<1$. 
Therefore, summarizing here the results of the following two sections,  we can conclude with the following statement for the adaptability of human social interactions

\begin{framed}
\hspace{-.25in} {\bf The adaptability of human social interactions}\\
The adaptability of human social interactions to technology can be seen as an effective way to modulate the parameter $\beta$ in Eq. $(\ref{f})$ parametrizing the probability to start or to end the social interactions.
\end{framed}

\subsection{Model of face-to-face interactions}
Here we recall the model of face-to-face interactions presented in   \cite{Stehle:2010,Zhao:2011} and we delineate the main characteristics and outcomes.
 A simple stochastic dynamics is imposed to the agent-based system in order to  model  face-to-face interactions. Starting from given initial conditions, the dynamics of face-to-face interactions at each time step $t$ is implemented as the following algorithm. 

\begin{itemize}
\item[(1)] An agent $i$ is chosen randomly.
\item[(2)] The agent $i$ updates his/her state $n_i=n$ with probability $f_n(t,t_i)$.

If the state $n_i$ is updated, the subsequent action of the agent proceeds with the following rules.
\begin{itemize}
\item[(i)] If  the agent $i$ is non-interacting ($n_i=1$), he/she starts an interaction with another non-interacting agent $j$ chosen with probability proportional to  $f_1(t, t_j)$. Therefore the coordination number of the agent $i$ and of the agent $j$ are updated ($n_i \rightarrow 2$ and $n_j \rightarrow 2$).
\item[(ii)] If the agent $i$ is interacting in a group ($n_i=n>1$), with probability $\lambda$ the agent leaves the group and with probability $1-\lambda$ he/she introduces an non-interacting agent to the group.
If the agent $i$  leaves the group,  his/her coordination number is updated ($n_i \rightarrow 1$) and also the coordination numbers of all the agents in the original group are updated  ($n_r \rightarrow n-1$, where $r$ represent a generic  agent in the original group). On the contrary, if the agent $i$  introduces another isolated agent $j$ to the group, the agent $j$ is chosen with probability proportional to  $f_1(t,t_j)$ and the coordination numbers of all the interacting agents are updated ($n_i \rightarrow n+1$, $n_j \rightarrow n+1$ and $n_r \rightarrow n+1$ where  $r$ represents  a  generic agent in the group ). 
\end{itemize}
\item[(3)] Time $t$ is updated as $t \rightarrow t+1/N$ (initially $t=0$). The algorithm is repeated from (1) until $t=T_{max}$. 
\end{itemize}
We have taken in the reinforcement dynamics with parameter $\beta=1$ such that
\begin{equation}
f_n(t, t')=\frac{b_n}{1+(t-t')/N}.
\label{p}
\end{equation}
 In Eq. $(\ref{p})$,for simplicity,  we take  $b_n=b_2$ for every $n\geq2$, indicating the fact the interacting agents change their state independently on the coordination number $n$.

We note that in this model we assume that everybody can interact with everybody so that the underline network model is fully connected. This seems to be a  very reasonable assumption if we want to model face-to-face interactions in small conferences,  which are venues designed to stimulate interactions between the participants. Nevertheless the model can be easily modified by embedding the agents in a social network so that interactions occur only between social acquaintances.

In the following we review the mean-field solution to this model. For the detailed description of the solution of the outline non-equilibrium dynamics the interest reader can see   \cite{Stehle:2010, Zhao:2011}. We denote by $N_n(t,t')$ the number of
agents interacting with  $n=0,1,\ldots,N-1$ agents at time $t$, who
have not changed state since time $t'$. In the mean field
approximation, the evolution equations for $N_n(t,t')$ are given by
  
\bea
\frac{\partial N_1(t,t')}{\partial t}&=&-2\frac{N_1(t,t')}{N}f_1(t-t')-(1-\lambda)\epsilon(t) \frac{N_1(t,t')}{N}f_1(t-t')+\sum_{i > 1}\pi_{i,2}(t)\delta_{tt'}\nonumber\\
\frac{\partial N_2(t,t')}{\partial t}&=&-2\frac{N_2(t,t')}{N}f_2(t-t')+[\pi_{1,2}(t)+\pi_{3,2}(t)]\delta_{tt'} \nonumber \\
\frac{\partial N_n(t,t')}{\partial t}&=&-n \frac{N_n(t,t')}{N}f_n(t-t') +[\pi_{n-1,n}(t)+\pi_{n+1,n}(t)+\pi_{1,n}(t)]\delta_{tt'},~n> 2. 
\label{dNiB}
\eea

In these equations, the parameter $\epsilon(t)$ indicates the rate
at which isolated nodes are introduced by another agent in already
existing groups of interacting agents. Moreover, $\pi_{mn}(t)$
indicates the transition rate at which agents change its state 
 from $m$ to $n$ (i.e. $m\to n$) at time $t$. In the mean-field
approximation the value of $\epsilon(t)$ can be expressed in terms of
$N_n(t,t')$ as
\be \epsilon(t)=\frac{\sum_{n >
    1}\sum_{t'=1}^tN_n(t,t')f_n(t-t')}{\sum_{t'=1}^tN_1(t,t')f_1(t-t')}.
\label{epsilon}
\ee
Assuming that asymptotically in time $\epsilon(t)$ converges to a time
independent variable, i.e. $\lim_{t\to
  \infty}\epsilon(t)=\hat{\epsilon}$, the solution to the rate
equations (\ref{dNiB}) in the large time limit is given by
\bea
\label{NiB}
N_1(t,t')&=&N_1(t',t')\bigg(1+\frac{t-t'}{N}\bigg)^{-b_1[2+(1-\lambda)\hat{\epsilon}]} \nonumber\\
N_2(t,t')&=&N_2(t',t')\bigg(1+\frac{t-t'}{N}\bigg)^{-2b_2} \\
N_n(t,t')&=&N_n(t',t')\bigg(1+\frac{t-t'}{N}\bigg)^{-nb_2}  \ \mbox{for} \ n> 2,\nonumber
\eea
with
\bea
N_1(t',t')&=&\sum_{n > 1}\pi_{n,1}(t')\nonumber \\
N_2(t',t')&=&\pi_{1,2}(t')+\pi_{3,2}(t')  \\
N_n(t',t')&=&\pi_{n-1,n}(t')+\pi_{n+1,n}(t')+\pi_{0,n}(t') \ \mbox{for} \ n> 2.\nonumber
\label{pig}.
\eea

\begin{figure}
\begin{center}
\includegraphics[width=0.5\textwidth]{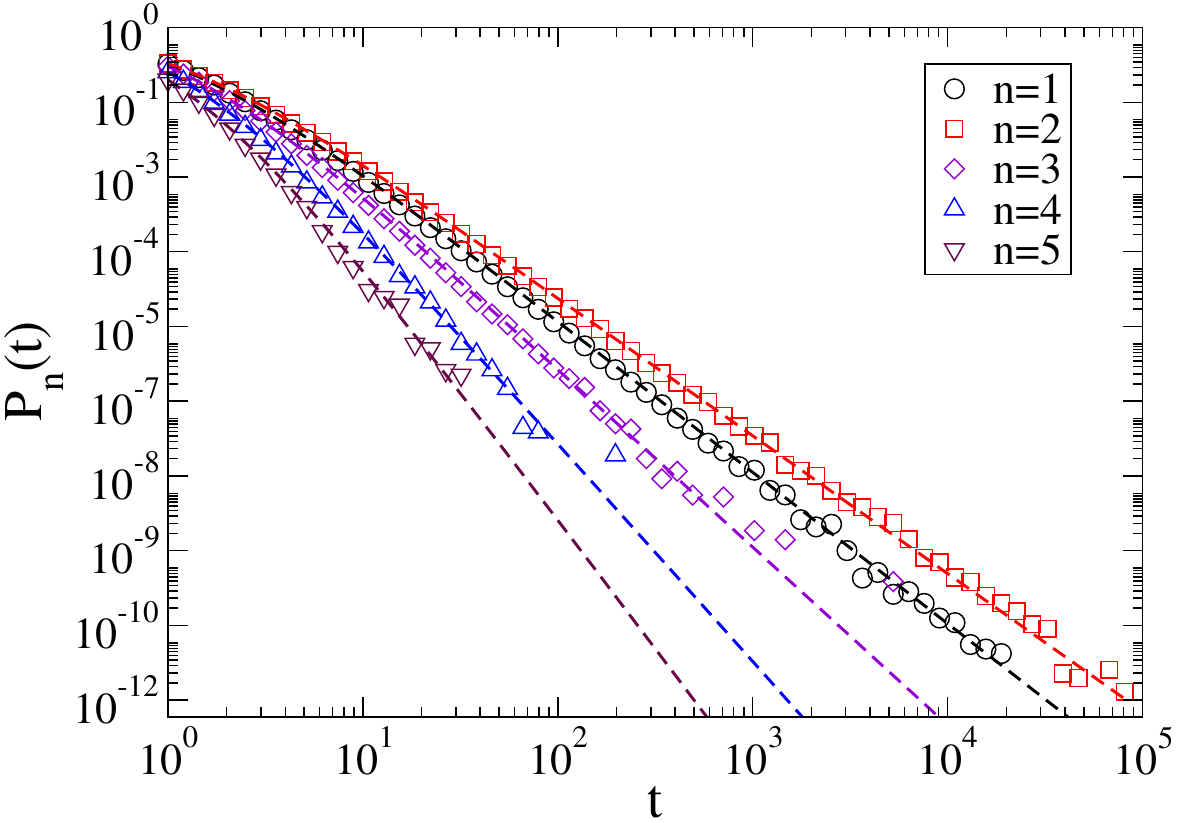}
  \end{center}
\caption{ \small{Distribution $P_n(\tau) $ of durations of
  groups of size $n$ in the stationary region. The simulation is
  performed with $N=1000$ agents for a number of time steps
  $T_{max}=N\times 10^5$. The parameter used are $b_0=b_1=0.7$,
  $\lambda=0.8$. The data is averaged over $10$ realizations.}}

\label{Groups_stationary}
\end{figure}

We denote by $P_n(\tau)$ the distribution of duration of different coordination number $n$ which satisfies the relation
\be
P_n(\tau)=\int_{t'=0}^{t-\tau}f_n(t-t')N(t,t')dt'.
\ee
and using Eq.(\ref{p}) and Eqs.(\ref{NiB}) we find that $P_n(\tau)$ simply satisfy
\bea
P_1(\tau) &\propto& (1+\tau)^{-b_1[2+(1-\lambda)\hat{\epsilon}]-1} \nonumber \\
P_n(\tau) &\propto& (1+\tau)^{-nb_2-1}.
\label{Pn}
\eea 
As shown in Fig.\ref{Groups_stationary}, the analytic prediction Eqs.(\ref{Pn}) is in good agreement with the computer simulation. 
\begin{figure}[h!]
\begin{center}
$\begin{array}{ccc}
\includegraphics[width=30mm, height=30mm]{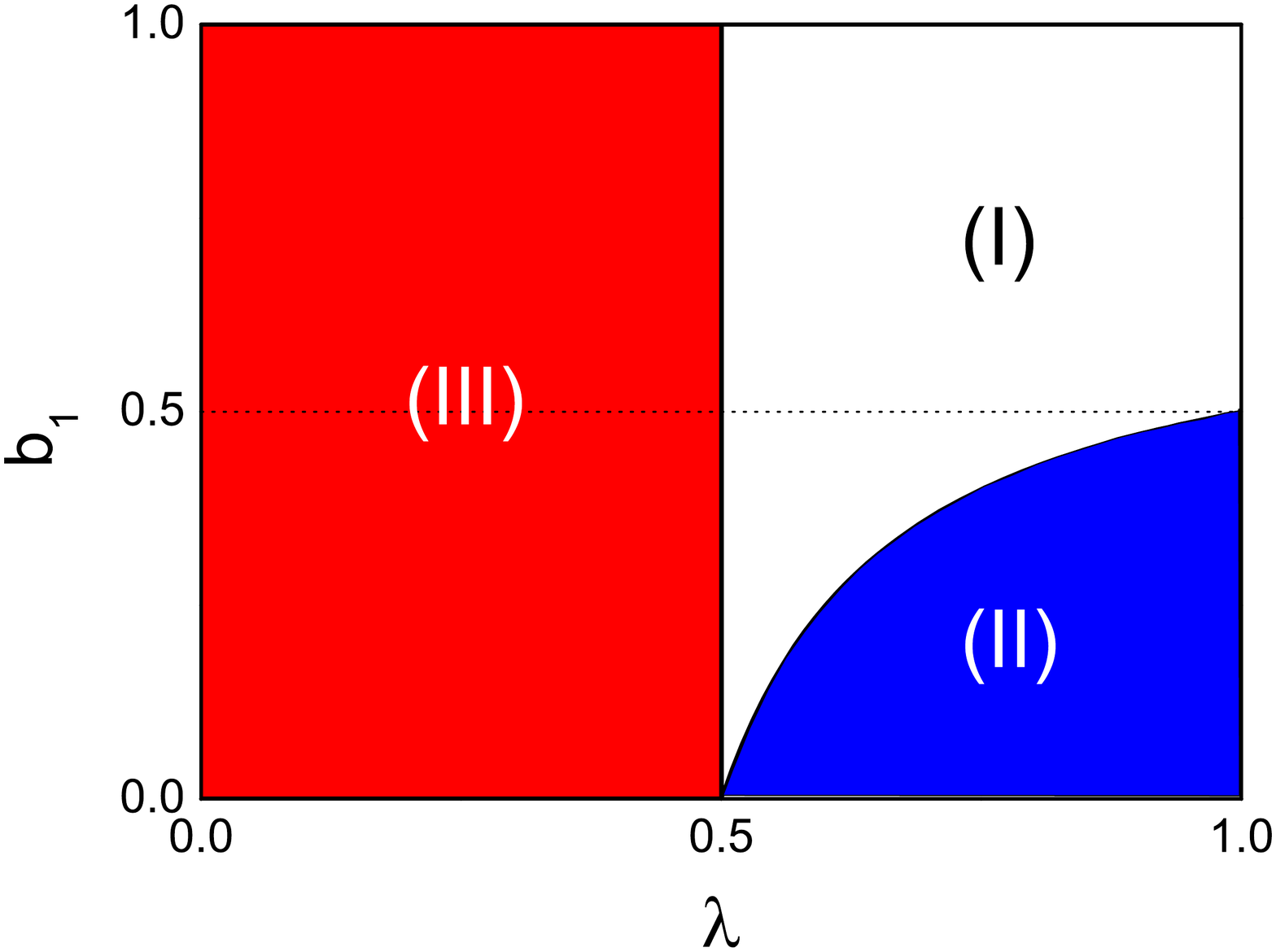}&
\includegraphics[width=30mm, height=30mm]{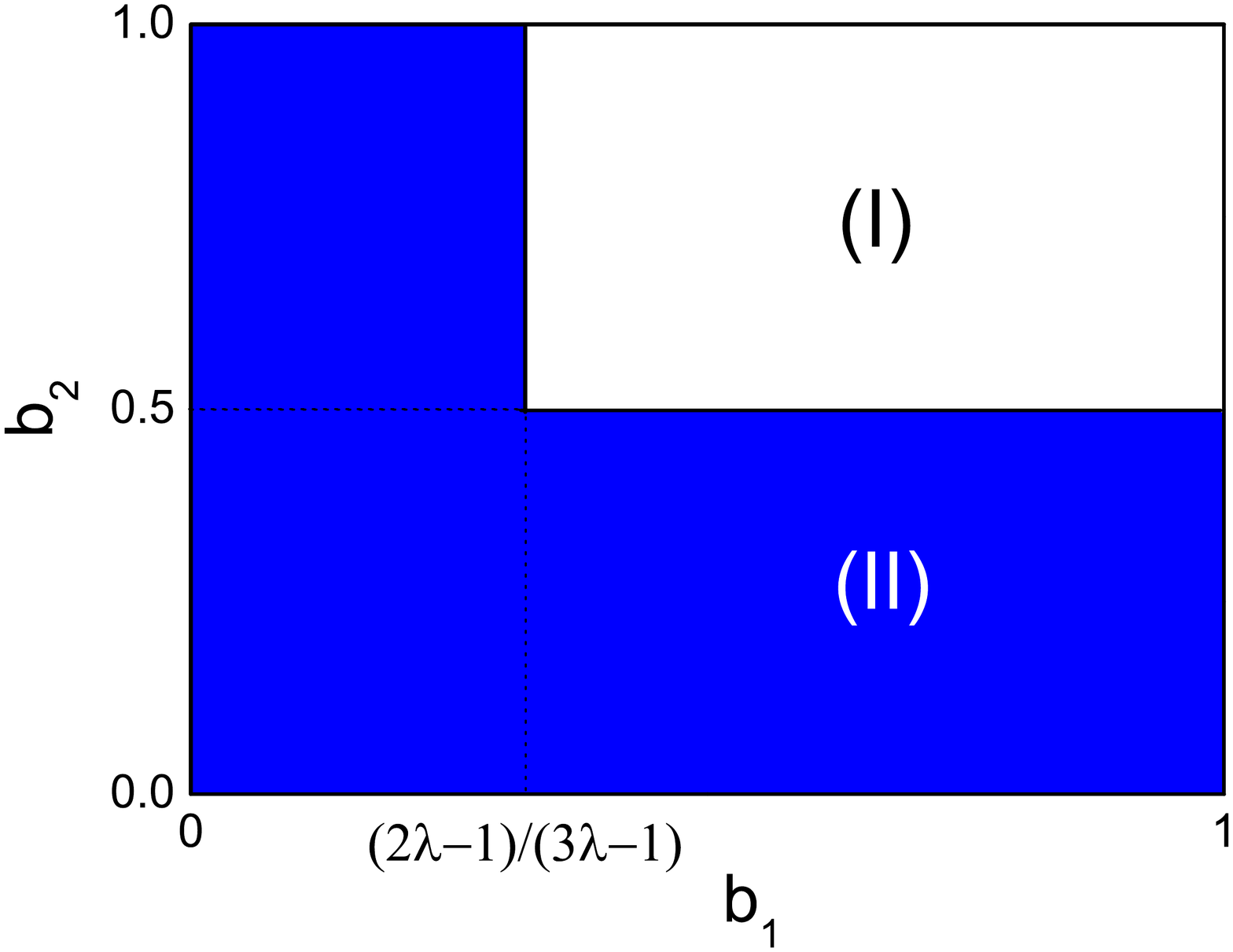}&
\includegraphics[width=30mm, height=30mm]{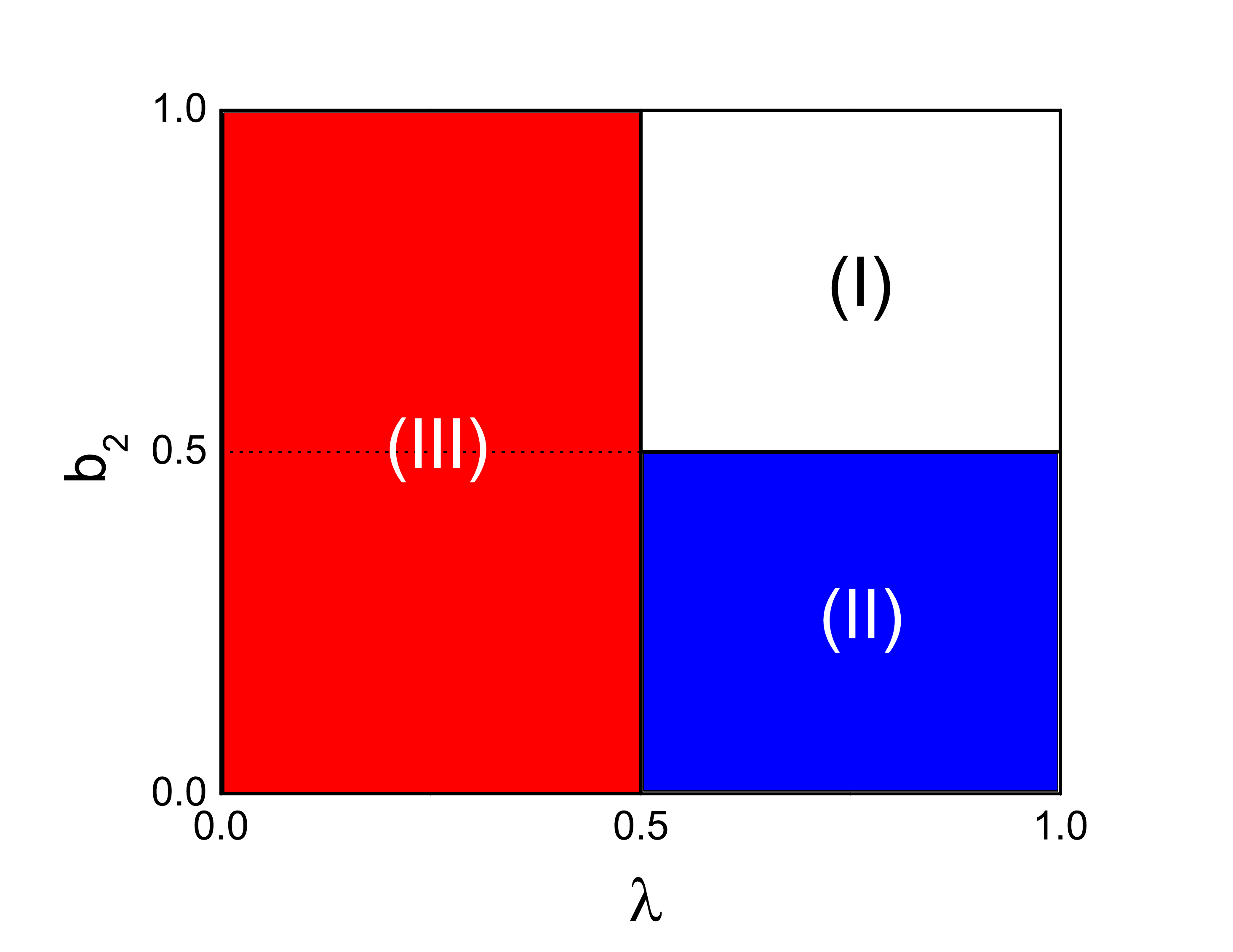}
\end{array}$
\end{center}
\caption{\small{Phase diagram of arbitrary state number $n$: The red area indicates the regime where a large group is formed and the solution is divergent. The blue area indicates the non-stationary regime. The white area indicates the stationary regime. }}
\label{Fig_phase}
\end{figure}

Despite the simplicity of this model, the non-equilibrium dynamics of this system is characterized by a non trivial phase diagram.
The phase-diagram of the model is summarized in Fig.\ref{Fig_phase}. We can distinguish between three phases:
\begin{itemize}
\item{\em Region I - the stationary region: $b_2>0.5$, $b_1>(2\lambda-1)/(3\lambda-3)$ and $\lambda>0.5$-} This region corresponds to the white  area in Fig.\ref{Fig_phase}. The region is stationary and the transition rates between different states are constant. 
\item{\em Region II - the non-stationary region: $b_2<0.5$ or $b_1>(2\lambda-1)/(3\lambda-3)$, and $\lambda>0.5$ -}This region corresponds to the blue area in Fig.\ref{Fig_phase}. The region is non-stationary and the transition rates between different states are decaying with time as power-law. 
\item {\em Region III - formation of a big group: $\lambda<0.5$ -}In this region there is an instability for the formation of a large group of  size  ${\cal O}(N)$. 
\end{itemize}

In both regions I and region II the distribution of the duration of groups of size $n$ follows a power-law distribution with an exponent which grows with the group size $n$.
This fact is well reproduced in the face-to-face data   \cite{Zhao:2011} and implies the following principle on the stability of groups in face-to-face interactions.

\begin{framed}
\hspace{-.25in} {\bf Stability of  groups in face-to-face interactions}
In face-to-face interactions, groups of larger size are less stable than groups of smaller size.
In fact the stability of a group depends on the independent decisions of the $n$ agents in the group to remain in contact.
\end{framed}

\subsection{Model of phone-call communication}
\label{sec3.2}
To model  cell-phone communication, we  consider once again a system of $N$ agents  representing the mobile phone users. Moreover, we introduce a static weighted network $G$, of which the nodes are  the agents in the system, the edges represent the social ties between the agents, such as friendships, collaborations or acquaintances, and the weights of the edges indicate the strengths of the social ties. Therefore the interactions between agents can only take place along the network $G$ (an agent can only interact with his/her neighbors on the network $G$). 
Here we propose a model for mobile-phone communication constructed with the use of the reinforcement dynamic mechanism. This model shares significant similarities with the previously discussed model for face-to-face interactions, but has two major differences. Firstly,   only pairwise interactions are allowed in the case of cell-phone communication. Therefore,  the state  $n$ of an agent only takes the values of either $1$ (non-interacting) or $2$ (interacting). Secondly, the probability that an agent ends his/her interaction depends on  the weight of  network $G$. The dynamics of cell-phone communication at each time step $t$ is then implemented as the following algorithm. 
\begin{itemize}
\item[(1)] An agent $i$ is chosen randomly at time $t$.
\item[(2)] 
The subsequent action of agent $i$  depends on his/her current state (i.e. $n_i$):
\begin{itemize}
\item[(i)] If $n_i=1$, he/she starts an interaction with one of his/her non-interacting neighbors $j$ of $G$ with probability $f_1(t_i,t)$ where $t_i$ denotes the last time at which agent $i$ has changed his/her state. If the interaction is started, agent $j$ is chosen randomly with probability proportional to $f_1(t_j,t)$ and the coordination numbers of agent $i$ and $j$ are then updated ($n_i \rightarrow 2$ and $n_j \rightarrow 2$).
\item[(ii)] If $n_i=2$, he/she ends  his/her current interaction with probability $f_2(t_i,t|w_{ij})$ where $w_{ij}$ is the weight of the edge between $i$ and the neighbor $j$ 
that is interacting with $i$. If the interaction is ended, the coordination numbers of agent $i$ and $j$ are then updated ($n_i \rightarrow 1$ and $n_j \rightarrow 1$).
\end{itemize}
\item[(3)] Time $t$ is updated as $t \rightarrow t+1/N$ (initially $t=0$). The algorithm is repeated from (1) until $t=T_{max}$. 
\end{itemize}
Here we take the probabilities $f_1(t,t^{\prime}), f_2(t,t^{\prime}|w)$ according to the following functional dependence
\bea
f_1(t,t')&=&f_1(\tau)=\frac{b_1}{(1+\tau)^{\beta}}\nonumber \\
f_2(t,t'|w)&=&f_2(\tau|w)=\frac{b_2g(w)}{(1+\tau)^{\beta}}
\label{f2t}
\eea
where the  parameters are chosen in the range $b_1>0$, $b_2>0$, $0 \le\beta\le 1$, $g(w)$ is a positive decreasing function of its argument, and $\tau$ is given by   $\tau=(t-t')/N$.

In order to solve the model analytically, we assume the quenched network $G$ to be annealed and uncorrelated. Here we outline the main result of this approach and we suggest for the interested reader to look at   papers   \cite{PlosOne, Frontiers} for the details of the calculations. Therefore we assume that the network is rewired while the degree distribution $p(k)$ and the weight distribution $p(w)$  remain constant. We denote by $N_1^k(t,t')$ the number of non-interacting agents with degree $k$ at time $t$ who have not changed their  state since time $t'$. Similarly we  denote by $N_2^{k,k',w}(t,t')$ the number of interacting agent pairs  (with degree respectively $k$ and $k'$ and weight of the edge $w$) at time $t$ who have not changed their states since time $t'$. In the annealed approximation the probability that an agent with degree $k$ is called by another agent  is proportional to its degree. Therefore the evolution equations of the model are given by 
\bea   
\frac{\partial N_1^k(t,t')}{\partial t}&=&-\frac{N_1^k(t,t')}{N}f_1(t-t')-ck\frac{N_1^k(t,t')}{N}f_1(t-t')+\pi_{21}^k(t)\delta_{tt'} \nonumber\\
\frac{\partial N_2^{k,k',w}(t,t')}{\partial t}&=&-2\frac{N_2^{k,k',w}(t,t')}{N}f_2(t-t'|w)+\pi_{12}^{k,k',w}(t)\delta_{tt'}
\label{dN1w}
\eea
where the constant $c$ is given by 
\be
c=\frac{\sum_{k'}\int_0^t dt' N_1^{k'}(t,t')f_1(t-t')}{\sum_{k'}k'\int_0^t dt'N^{k'}_1(t,t')f_1(t-t')}.
\label{c_sum}
\ee
 In  Eqs. $(\ref{dN1w})$ the rates $\pi_{pq}(t)$  indicate the average number of agents changing from state $p=1,2$ to state $q=1,2$ at time $t$.
 The solution of the dynamics must of course satisfy the conservation equation
 \be 
\int dt' \big[N_1^k(t,t')+\sum_{k',w} N_2^{k,k',w}(t,t')\big]=Np(k).
\label{N_conserve} 
\ee
In the following we will  denote by $P^k_1(t,t')$  the probability distribution that an agent with degree $k$ is  non-interacting in the period between  time $t'$ and time  $t$  and we will denote by $P^w_2(t,t')$  the probability that an interaction of weight $w$ is lasting from time $t'$ to time $t$ which satisfy
\bea
P_1^k(t,t')&=&(1+ck)f_1(t,t')N_1^k(t,t')\nonumber \\
P_2^w(t,t')&=&2f_2(t,t'|w)\sum_{k,k'}N_2^{k,k',w}(t,t'). 
\label{P12}
\eea
As a function of the value of the parameter of the model we found different distribution of duration of contacts and non-interaction times.

\begin{itemize}
\item {\em Case $0<\beta<1$}
The system allows always for  a stationary solution with $N_1^k (t,t')=N_1^k(\tau)$ and $N_2^{k,k',w}(t,t')=N_2^{k,k',w}(\tau)$.  
The distribution of duration of non-interaction times $P_1^k(\tau)$ for agents of degree $k$ in the network  and the distribution of interaction times  $P_2^w(\tau)$  for links of weight $w$ is given by 

\bea
P_1^k(\tau)&\propto &\frac{b_1(1+ck)}{(1+\tau)^{\beta}}e^{-\frac{b_1(1+ck)}{1-\beta}(1+\tau)^{1-\beta}}\nonumber \\
P_2^w(\tau)&\propto &\frac{2b_2g(w)}{(1+\tau)^{\beta}}e^{-\frac{2b_2g(w)}{1-\beta}(1+\tau)^{1-\beta}}.
\label{P2kt1}
\eea
 Rescaling Eqs.(\ref{P2kt1}), we obtain the Weibull distribution which is in good agreement with the results observed in mobile-phone datasets.
 \item {\em Case $\beta=1$}
 Another interesting limiting case of the mobile-phone communication model is the case $\beta=1$ such that $f_1^k(\tau)\propto(1+\tau)^{-1}$ and $f_2^w(\tau|w)\propto(1+\tau)^{-1}$. 
In this case the model is much similar to the model used to mimic face-to-face interactions described in the previous subsection   \cite{Stehle:2010,Zhao:2011}, but the interactions are binary and they occur on a weighted network. In this case we get the solution 
\bea
 N_1^k(\tau)& = & N\pi_{21}^k(1+\tau)^{-b_1(1+ck)}\nonumber \\
 N_2^{k,k',w}(\tau) & = & N\pi_{12}^{k,k',w}(1+\tau)^{-2b_2g(w)}.
\eea
and consequently the distributions of duration of given states Eqs. $(\ref{P12})$ are given by 
\bea
 P_1^k(\tau) & \propto & \pi_{21}^k(1+\tau)^{-b_1(1+ck)-1}\nonumber \\
 P_2^w(\tau) & \propto & \pi_{12}^{k,k',w}(1+\tau)^{-2b_2g(w)-1}.
\eea
The probability distributions are power-laws.This result remains  valid for every value of the parameters $b_1,b_2,g(w)$  nevertheless the stationary condition is only valid for  
\bea
b_1(1+ck)>1\nonumber \\
2b_2g(w)>1.
\eea
Indeed this condition ensures that the self-consistent constraints Eqs. (\ref{c_sum}),  and the conservation law Eq. (\ref{N_conserve}) have a stationary solution.

 \item {\em Case $\beta=0$}
 This is the case in which the process described by the model is a Poisson process and their is no reinforcement dynamics in the system.
 Therefore we find that the distribution of durations are exponentially distributed.
  In fact for $\beta =0$ the functions $f_1(\tau)$ and $f_2(\tau|w)$ given by Eqs.$(\ref{f2t})$ reduce to constants, therefore the process of creation of an interaction is a Poisson process. In this case the social interactions do not follow the  reinforcement dynamics. The solution  that we get for the number of  non interacting agents of degree $k$, $N_1^k(\tau)$ and the number of interacting pairs $N_2^{k,k'w}(\tau)$ is given by 
\bea
 N_1^k(\tau)&=&N\pi_{21}^ke^{-b_1(1+ck)\tau}\nonumber \\
 N_2^{k,k',w}(\tau)&=&N\pi_{12}^{k,k',w}e^{-2b_2g(w)\tau}.
\eea
Consequently the distributions of duration of given states Eqs. $(\ref{P12})$ are given by 
\bea
 P_1^k(\tau) \propto e^{-b_1(1+ck)\tau}\nonumber \\
P_2^w(\tau) \propto e^{-2b_2g(w)\tau}.
\eea
Therefore the  probability distributions $P_1^k(\tau)$  and $P_2^w(\tau)$  are  exponentials as expected in a Poisson process.

\end{itemize}

\begin{figure}[h!]
\begin{center}
\includegraphics[width=3.27in]{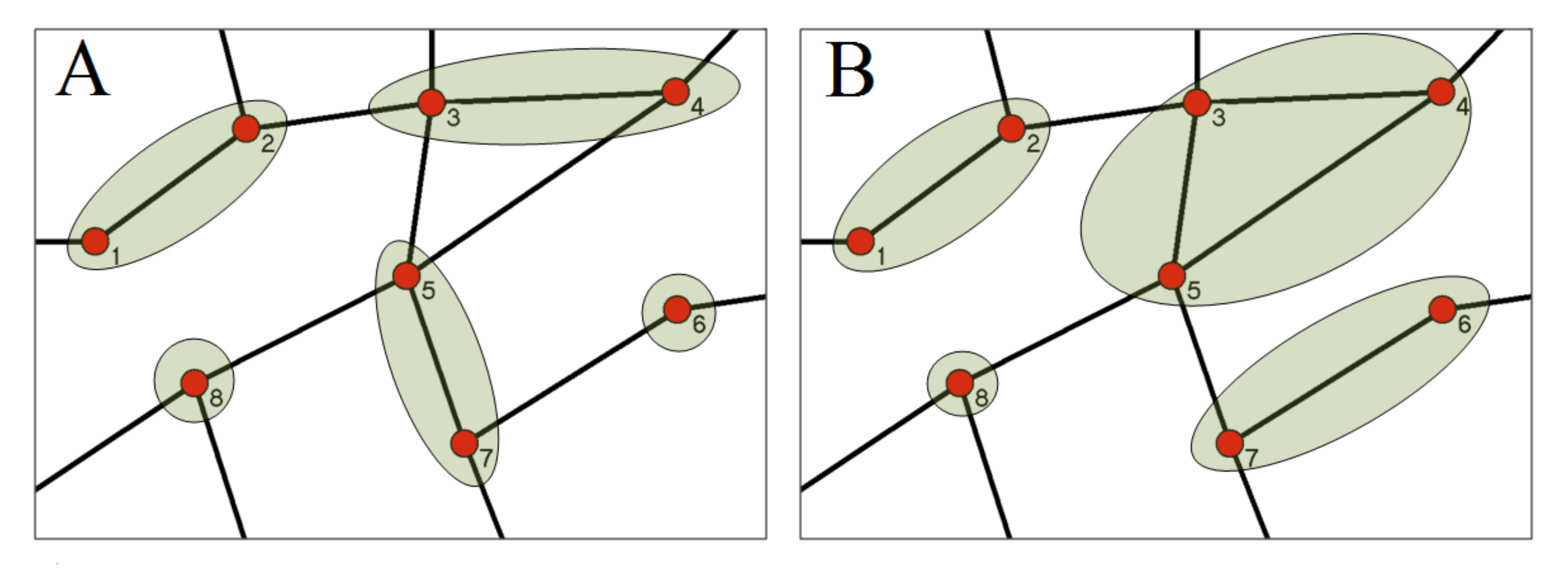}
\end{center}
\caption{  \small{The dynamical social networks are composed by different dynamically changing groups of interacting agents. (A) Only groups of size one or two are allowed as in the phone-call communication. (B) Groups of any size are allowed as in the face-to-face interactions. }}
\label{fig1}
\end{figure}

\begin{figure}[h!]
\begin{center}
\includegraphics[width=3.27in]{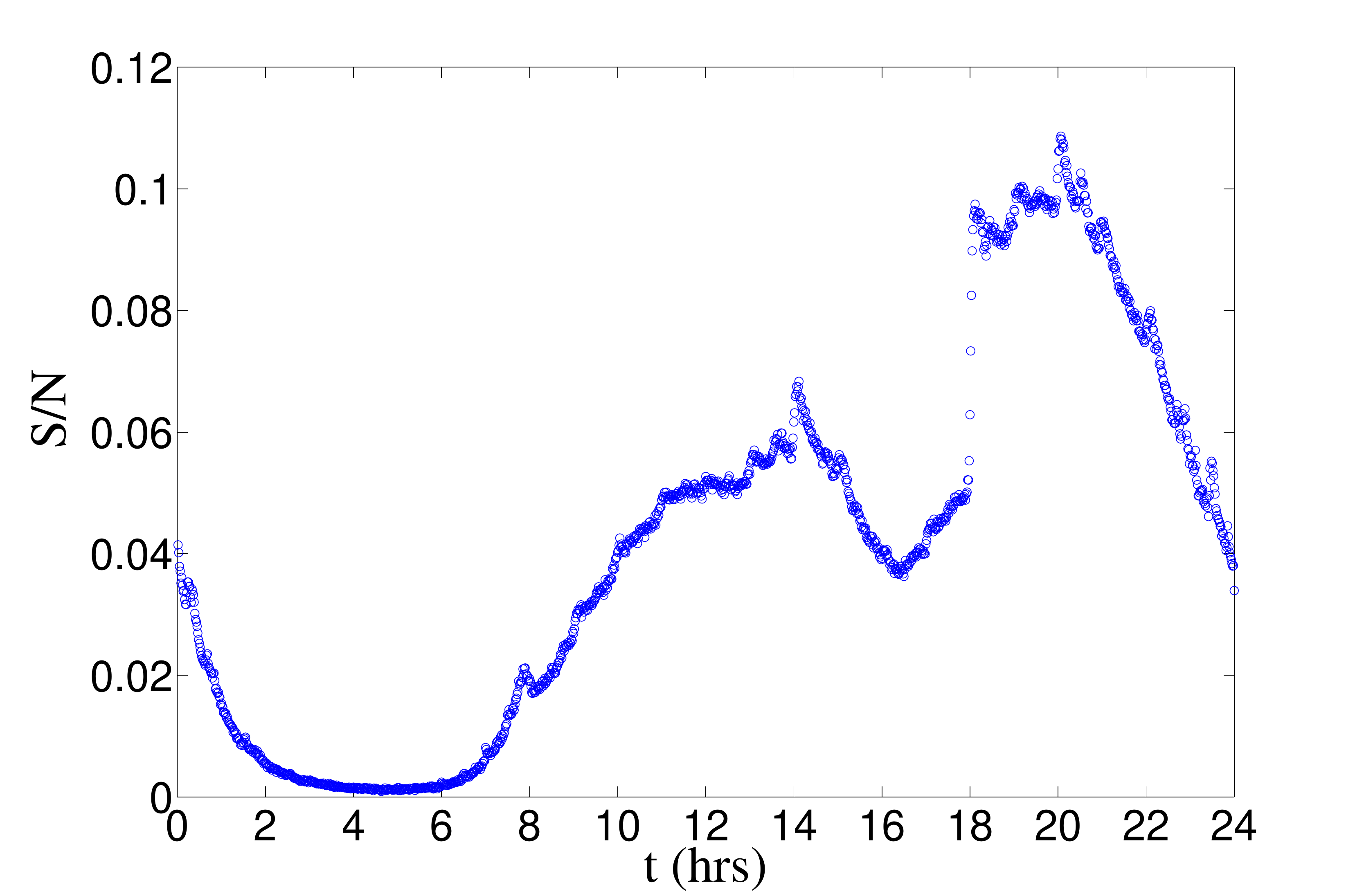}
\end{center}
\caption{  \small{Evaluation of the entropy of the dynamical social networks of phone calls communication in a typical week-day. In the nights the social dynamical network is more predictable. Figure from   \cite{PlosOne}.}}
\label{entropyt}
\end{figure}

\section{Entropy of temporal social networks}

In this section we introduce the entropy of temporal social networks as a measure of information encoded in their dynamics.  We can assume that the following  stochastic dynamics  takes place in the network: according to this dynamics at each time step $t$, different  interacting groups can be formed and can be dissolved giving rise to the temporal  social networks. The agents are embedded in a social network $G$ such that interaction can occur only by acquaintances between first neighbors of the network $G$. This is a good approximation if we want to model social interactions on the fast time scale. In the case of a small conference, where each participant is likely to discuss with any other participant we can consider a fully connected network as the underlying network $G$ of social interactions. In the network $G$   each set of  interacting agents  can be seen as a  connected subgraph of ${\cal G}$, as shown  in Fig \ref{fig1}.
We use an indicator function  $g_{i_1,i_2,\ldots , i_n}(t)$ to denote, at time $t$, the maximal set  $i_1$, $i_2$,..., $i_n$ of interacting agents in a group. If $(i_1i_2,\ldots, i_n)$ is the maximal set of interacting agents in a group,  we let $g_{i_1,i_2,\ldots , i_n}(t)=1$ otherwise we put $g_{i_1,i_2,\ldots, i_n}(t)=0$. Therefore at any given time the following relation is satisfied, 
\begin{equation}
\sum_{{\cal G}=(i,i_2,\ldots,i_n )|i\in {\cal G}}g_{i,i_2,\ldots, i_n}(t)=1.
\end{equation} 
where ${\cal G}$ is  an arbitrary connected subgraph of $G$.
Then we denote by ${\cal S}_t=\{g_{i_1,i_2,\ldots, i_n}(t^{\prime})\, \forall t^{\prime}<t\}$ the history of the dynamical social networks, and $p(g_{i,i_2,\ldots, i_n}(t)=1|{\cal S}_t)$ the probability that $g_{i_1,i_2,\ldots, i_n}(t)=1$ given the history ${\cal S}_t$. 
Therefore the likelihood that at time $t$ the dynamical social networks has a group configuration $g_{i_1,i_2,\ldots,i_n}(t)$ is given by 
\begin{equation}
{\cal L}=\prod_{{\cal G}} p(g_{i_1,i_2,\ldots, i_n}(t)=1|{\cal S}_t)^{g_{i_1,i_2,\ldots, i_{n}}(t)}.
\label{Likelihood}
\end{equation}

We denote the entropy of the dynamical networks as $S=-\Avg{\log{\cal L}}_{|{\cal S}_t}$ indicating the logarithm of the typical number of all possible group configurations at time $t$ which can be explicitly written as
\begin{equation}
S=- \sum_{{\cal G}}p(g_{i,i_2,\ldots, i_n}(t)=1|{\cal S}_t)\log p(g_{i,i_2,\ldots, i_n}(t)=1|{\cal S}_t).
\end{equation}

The value of the entropy can be interpreted as following: if the entropy is larger, the dynamical network is less predictable, and   several possible dynamic configurations of groups are expected in the system at time $t$. On the other hand, a smaller entropy indicates a smaller number of possible future configuration and a temporal network state which is more predictable. 

\subsection{Entropy of phone-call communication}
In this subsection we discuss the evaluation of the entropy of phone-call communication. For phone-call communication, we only allow pairwise interaction in the system such that the product in Eq.(\ref{Likelihood}) is only taken over all single nodes and edges of the quenched network $G$ which yields
\begin{equation}
{\cal L}=\prod_i p(g_i(t)=1|{\cal S}_t)^{g_i(t)}\prod_{ij|a_{ij}=1} p(g_{ij}(t)=1|{\cal S}_t)^{g_{ij}(t)} 
\end{equation}
with 
\begin{equation}
g_i(t)+\sum_{j} a_{ij} g_{ij}(t)=1.
\end{equation}
where $a_{ij}$ is the adjacency matrix of $G$.
The entropy then takes a simple form
\begin{eqnarray}
S&=&- \sum_i p(g_{i}(t)=1|{\cal S}_t)\log p(g_{i}(t)=1|{\cal S}_t)\nonumber \\
&&-\sum_{ij}a_{ij} p(g_{ij}(t)=1|{\cal S}_t)\log p(g_{ij}(t)=1|{\cal S}_t).
\label{s_pair2}
\end{eqnarray}


%
\subsection{Analysis of the entropy of a large dataset of mobile phone communication}
\label{sec:3}
In this subsection we use the entropy of temporal social networks to analyze the information encoded in a major European mobile service provider, making use of the same dataset that we have  used to measure the distribution of call duration in Section 2. Here we evaluate   the entropy of the temporal networks formed by the phone-call communication in a typical week-day in order to study how the entropy of temporal social networks is affected by circadian rhythms of human behavior. 

For the evaluation of the entropy of temporal social networks we consider a subset of the large dataset of mobile-phone communication. We selected $562,337$ users who executed at least one call  a day during a weeklong period. We denote by $f_n(t,t^{\prime})$ the transition probability that an agent  in state $n$ ($n=1,2)$ changes its state at time $t$ given that he/she has been in his/her current state for a duration $\tau=t-t^{\prime}$. The probability $f_n(t,t^{\prime})$ can be estimated directly from the data. Therefore, we  evaluate the  entropy in a typical weekday of the dataset by using the transition probabilities $f_n(t,t^{\prime})$ and the definition of entropy of temporal social networks (Readers should refer to the supplementary material of Ref.   \cite{PlosOne} for the details). In Fig. \ref{entropyt} we show the resulting evaluation of entropy  in a typical day of our  phone-call communication dataset. The entropy of the temporal social network is plotted as a function of time during one typical day. The mentioned figure shows evidence that the entropy of temporal social networks changes significantly during the day  reflecting the circadian rhythms of human behavior. 

\subsection{Entropy modulated by the adaptability of human behavior}
The adaptability of human behavior is evident when comparing the distribution ofthe  duration of phone-calls with the duration of face-to-face interactions.
In  the framework of the model for mobile-phone interactions described in Sec. \ref{sec3.2}, this adaptability, can be understood, as a possibility to change the exponent $\beta$ in Eqs. (\ref{f}) and (\ref{f2t}) regulating the duration of social interactions.

\begin{figure}[h!]
\begin{center}
\includegraphics[width=3.27in ,height=50mm]{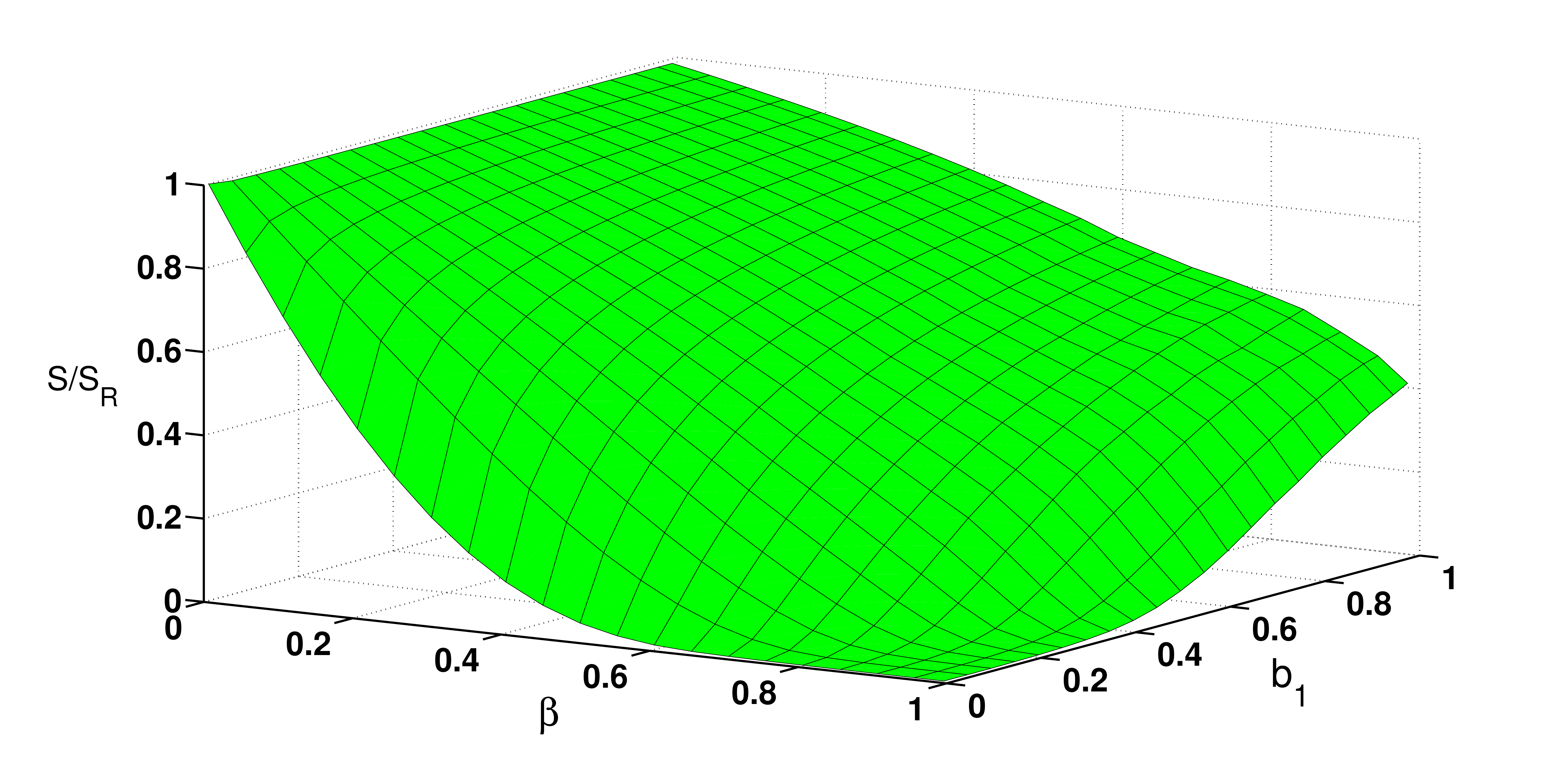}
\end{center}
\caption{  \small{Entropy $S$ of social dynamical network model of pairwise communication normalized with the entropy $S_R$ of a null model in which the expected average duration of phone-calls is the same but the distribution of duration of phone-calls and non-interaction time are Poisson distributed. The network size is $N=2000$ the degree distribution of the network is exponential with average $\avg{k}=6$, the weight distribution is $p(w)=Cw^{-2}$ and $g(w)$ is taken to be $g(w)=b_2/w$ with $b_2=0.05$.
The value of $S/S_R$ is depending on the two parameters $\beta, b_1$. For every value of $b_1$ the normalized entropy is smaller for $\beta\to 1$. Figure from   \cite{PlosOne}.}
}
\label{entropy_network}
\end{figure}

Changes in the parameter $\beta$ correspond to  different values entropy of the dynamical social networks. Therefore, by modulating the exponent $\beta$, the human behavior is able to modulate the information encoded in temporal social networks.
In order to show the effect on entropy of a variation of the exponent $\beta$ in the dynamics of social interaction networks, we considered the entropy corresponding to the model described in Sec. \ref{sec3.2} as  a function of the parameters $\beta$ and $b_1$ modulating the probabilities $f_1(t,t^{\prime}), f_2(t,t'|w)$ Eqs.(\ref{f2t}). 
In Fig.  \ref{entropy_network} we report the entropy $S$  of the proposed model  a function of $\beta$ and $b_1$. The entropy $S$, given by  Eq.(\ref{s_pair2}), is calculated using the  annealed approximation for the solution of the model and assuming the large network limit.  In the calculation of the entropy $S$ we have taken  a  network of size $N=2000$ with exponential degree distribution of average degree $\avg{k}=6$, weight distribution $P(w)=Cw^{-2}$ and function $g(w)=1/w$ and  $b_2=0.05$.  Our aim in   Fig. \ref{entropy_network} is to show  only the effects on the entropy due to the different distributions of duration of contacts and non-interaction periods. Therefore we have normalized the entropy $S$ with the entropy $S_R$ of  a null model of social interactions in which the duration of groups are Poisson distributed but the average time of interaction and non-interaction time are the same as in the model of cell-phone communication (Readers should refer to the supplementary material of Ref.   \cite{PlosOne} for more details). 
From Fig. \ref{entropy_network}  we observe that if we keep $b_1$ constant,  the ratio $S/S_R$ is a decreasing function of the parameter $\beta$. This  indicates that the broader is the distribution of probability of duration of contacts, the higher is the information encoded in the dynamics of the network. Therefore the heterogeneity in the distribution of duration of contacts and no-interaction periods implies higher level of information in the social network. The human adaptive behavior by changing the exponent $\beta$ in face-to-face interactions and mobile phone communication effectively changes the entropy of the dynamical network.

\section{Conclusions}
The goal of network science is to model, characterize, and predict the behavior of complex networks.
Here, in this chapter, we have delineated a first step in the characterization of the information encoded in temporal social networks. In particular we have focused on modelling phenomenologically social interactions on the fast time scale, such a face-to-face interactions and mobile phone communication activity. Moreover, we have defined the entropy of dynamical social networks, which is able to quantify the information present in social network dynamics.
We have found that human social interactions are bursty and adaptive. Indeed,  the duration of social contacts can be modulated by the adaptive behavior of humans: while in face-to-face interactions dataset a power-law distribution of duration of contacts has been observed, we have found, from the analysis of a large dataset of mobile-phone communication, that mobile-phone calls are distributed according to a Weibull distribution.
We have modeled this adaptive behavior by assuming that the dynamics underlying the formation of social contacts implements a reinforcement dynamics according to which the longer an agent has been in a state (interacting or non-interacting) the less likely it is that he will change his/her state.
We have used the  entropy of dynamical social networks to evaluate the information present in the temporal network of mobile-phone communication, during a typical weekday of activity,  showing that the information content encoded in the dynamics of the network changes during a typical day.
Moreover, we have compared  the entropy in a social network with the duration of contacts following a Weibull distribution, and with the duration of contacts following a power-law in the framework of the  stochastic model proposed for mobile-phone communication.
We have found that a modulation of the statistics of duration of contacts strongly modifies the information contents present in the dynamics of temporal networks.
Finally, we conclude that the  duration of social contacts in humans has a distribution that strongly deviates from an exponential. Moreover, {the} data show that human behavior is able to modify the information encoded in social networks dynamics during the day and when {facing} a new technology such as the mobile-phone communication technology. 

\subsection{Acknowledgement}
We thank A. Barrat and J. Stehl\'e for a collaboration that started our research on face-to-face interactions. Moreover we especially  thank A.-L. Barab\'asi for his useful comments and for the mobile call data used in this research.
MK acknowledges the financial support from EU’s 7th Framework Program’s FET-Open to ICTeCollective project no. 238597


\begin{thebibliography}{100}
\bibitem{Dorogovtsev:2003}
 Dorogovtsev SN, Mendes JFF (2003) 
{ Evolution of networks: From biological nets to the {I}nternet and {WWW}}.
Oxford Univ Press.


\bibitem{Newman:2003} 
Newman MEJ (2003) 
The structure and function of complex networks. 
{ SIAM Rev} 45:157–256.


\bibitem{Boccaletti:2006}
Boccaletti S, Latora V, Moreno Y, Chavez M, Hwang DU (2006) 
Complex networks: Structure and dynamics. 
{ Phys Rep} 424:175–308.

\bibitem{Caldarelli:2007} 
Caldarelli G (2007) 
{ Scale-Free Networks}. 
Oxford Univ Press.
\bibitem{Doro_review}
Dorogovtsev SN, Goltsev AV and Mendes JFF (2008)
Rev. Mod. Phys. 80:1275 

\bibitem{Barrat:2008} 
Barrat A, Barth\'elemy M, Vespignani A (2008) 
{ Dynamical processes on complex networks} 
(Cambridge Univ Press, Cambridge).  

\bibitem{Fortunato}
Castellano C, Fortunato S, Loreto V (2009) 
Statistical physics of social dynamics. 
{ Rev Mod Phys} 81:591-646.


\bibitem{Palla:2007}
Palla G, Barab\'asi AL, Vicsek T (2007) 
Quantifying social group evolution. 
{ Nature} 446:664-667. 


\bibitem{Lehmann}
Ahn YY, Bagrow JP, Lehmann S (2010)
Link communities reveal multiscale complexity in networks. 
{ Nature} 466:761-764. 


\bibitem{Bianconi:2009}
Bianconi G, Pin P, Marsili M (2009) 
Assessing the relevance of node features for network structure. 
{ Proc Natl Acad Sci USA} 106:11433-11438.
\bibitem{Ising}
Bianconi G (2002)
Phys. Lett. A 303:166

\bibitem{Ising_spatial}
Bradde S, Caccioli F, Dall'Asta L and Bianconi G (2010)
Phys. Rev. Lett. 104:218701 
\bibitem{Holme:2005}
Holme P (2005) 
Network reachability of real-world contact sequences. 
{ Phys Rev E} 71:046119.

\bibitem{Latora:2009}
Tang J, Scellato S, Musolesi M, Mascolo C, Latora V (2010) 
Small-world behavior in time-varying graphs. 
{ Phys Rev E} 81:055101.
\bibitem{Havlin:2009}
Parshani R, Dickison M, Cohen R, Stanley HE, Havlin S (2010) 
Dynamic networks and directed percolation. 
{ Europhys Lett} 90:38004.
\bibitem{Cattuto:2010} 
Cattuto C, Van den Broeck W, Barrat A, Colizza V, Pinton JF, Vespignani A (2010) 
Dynamics of person-to-person interactions from distributed RFID sensor networks. 
{ PLoS ONE} 5:e11596.
\bibitem{Isella:2011}
Isella L, Stehl\'e J, Barrat A, Cattuto C, Pinton JF, Van den Broeck W (2011) 
What's in a crowd? Analysis of face-to-face behavioral networks. 
{ J Theor Biol} 271:166-180.

\bibitem{Holme:2012}
 Holme P and  Saram\"aki J (2012)
 {Phys. Rep.} 519:97-125.


\bibitem{Granovetter:1973}
Granovetter M (1973) 
The strength in weak ties. 
{ Am J Sociol} 78:1360–1380.


\bibitem{Wasserman:1994}
Wasserman S, Faust K (1994) 
{ Social Network Analysis: Methods and applications}.
Cambridge Univ Press.





\bibitem{Eagle:2006}
Eagle N, Pentland AS (2006) 
Reality mining: sensing complex social systems. 
{ Personal Ubiquitous Comput} 10:255-268.



\bibitem{Hui:2005}
Hui P, Chaintreau A, Scott J, Gass R, Crowcroft J, Diot C (2005) 
Pocket switched networks and human mobility in conference environments. 
{ Proceedings of the 2005 ACM SIGCOMM workshop on Delay-tolerant networking} (Philadelphia, PA) pp 244-251.

\bibitem{Onnela:2007}
Onnela JP,  Saram\"aki J, Hyv\"onen J, Szab\'o G, Lazer D, Kaski K, Kert\'esz J,  Barab\`asi AL (2007) 
Structure and tie strengths in mobile communication networks. 
{ Proc Natl Acad Sci USA} 104:7332-7336.
\bibitem{Brockmann:2006} 
Brockmann D, Hufnagel L, Geisel T (2006) 
The scaling laws of human travel. 
{ Nature} 439:462-465.


\bibitem{Gonzalez:2008}
Gonz\'alez MC, Hidalgo AC, Barab\'asi AL (2008) 
Understanding individual human mobility patterns. 
{ Nature} 453:779-782.


\bibitem{Bornholdt:2002}
Davidsen J, Ebel H, Bornholdt S (2002) 
Emergence of a Small World from Local Interactions: Modeling Acquaintance Networks. 
{ Phys Rev Lett} 88:128701.
\bibitem{Marsili:2004}
Marsili M, Vega-Redondo F, Slanina F (2004) 
The rise and fall of a networked society: A formal model. 
{ Proc Natl Acad Sci USA} 101:1439-1442.
\bibitem{Holme:2006}
Holme P, Newman MEJ (2006) 
Nonequilibrium phase transition in the coevolution of networks and opinions. 
{ Phys Rev E} 74:056108.
(2006).
\bibitem{MaxiSanMiguel:2008}
Vazquez F, Egu\'iluz VM, San Miguel M (2008) 
Generic Absorbing Transition in Coevolution Dynamics. 
{ Phys Rev Lett} 100:108702.



\bibitem{Barabasi:2005}
Barab\'asi AL (2005) 
The origin of bursts and heavy tails in humans dynamics. 
{ Nature} 435:207-211.

\bibitem{Vazquez:2006}
V\'azquez A, et al. (2006)
Phys. Rev. E 73:036127.




\bibitem{Rybski:2009}
Rybski D, Buldyrev SV, Havlin S, Liljeros F, Makse HA (2009) 
Scaling laws of human interaction activity. 
{ Proc Natl Acad Sci USA} 106:12640-12645. 


\bibitem{Amaral:2009}
Malmgren RD, Stouffer DB, Campanharo ASLO, Nunes Amaral LA (2009) 
On universality in human correspondence activity. 
{ Science} 325:1696-1700.



\bibitem{Scherrer:2008}
Scherrer A, Borgnat P,  Fleury E,  Guillaume JL, Robardet C (2008) 
Description and simulation of dynamic mobility networks. 
{ Comp Net} 52:2842-2858.

\bibitem{Stehle:2010}
Stehl\'e J, Barrat A, Bianconi G (2010) 
Dynamical and bursty interactions in social networks. 
{ Phys Rev E} 81:035101.


\bibitem{Zhao:2011}
Zhao K, Stehl\'e J, Bianconi G, Barrat A (2011) 
Social network dynamics of face-to-face interactions. 
{ Phys Rev E} 83:056109.


\bibitem{Karsai:2012}
Karsai M, Kaski K, Barab\'asi AL, Kert\'esz J (2012)
Universal features of correlated bursty behaviour
{ Sci. Rep. } 2:397.

\bibitem{Karsai:2011b}
Jo HH, Karsai M, Kert\'esz J, Kaski K (2012) 
Circadian pattern and burstiness in mobile phone communication.
New Jour. Physics 14:013055



\bibitem{Vazquez:2007}
V\'azquez A, R\'acz B, Lukacs A, Barab\`asi AL (2007) 
Impact of Non-Poissonian activity patterns on spreading processes. 
{ Phys Rev Lett} 98:158702.


 
\bibitem{Karsai:2011a}
Karsai M, Kivel\"a M, Pan R K, Kaski K, Kert\'esz J, Barab\'asi A-L, Saram\"aki J (2011) 
Small but slow world: How network topology and burstiness slow down spreading
{ Phys Rev E} 83:025102.



\bibitem{PlosOne}
Zhao K, Karsai M and Bianconi G (2011)
Entropy of Dynamical Social Networks
PloSOne 6:e28116

\bibitem{Frontiers}
Zhao K and Bianconi G (2011)
Social Interaction Model  and adaptability of human behavior
Front.  Physiol. 2:101



\bibitem{Bisson}
Bisson G, Bianconi G, Torre V (2012)
the dynamics of group formation among leeches
Front.  Physio.  3:133

\bibitem{Chialvo}
Anteneodo C, Chialvo DR (2009) 
Unraveling the fluctuations of animal motor activity. 
{ Chaos} 19:033123.
 \bibitem{Motter}
Altmann EG, Pierrehumbert JB, Motter AE (2009) 
Beyond Word Frequency: Bursts, Lulls, and Scaling in the Temporal Distributions of Words. 
{ PLoS ONE} 4:e7678.
\bibitem{Lambiotte2}
Quercia D, Lambiotte R, Stillwell D, Kosinski M, Crowcroft J (2012)
The Personality of Popular Facebook Users,
ACM CSCW 12 :955-964.
\bibitem{Cover:2006}
Cover T and Thomas JA (2006)
{ Elements of Information Theory}.
Wiley-Interscience.


\bibitem{Kleinberg}
Kleinberg JM (2000) 
Navigation in a small world. 
{ Nature} 406:845.


\bibitem{WS}
Watts DJ, Strogatz SH (1998) 
Collective dynamics of ‘small-world’ networks. 
{ Nature} 393:440-442.
\bibitem{Newman:2001}
Newman MEJ (2001) 
The structure of scientific collaboration networks. 
{ Proc Natl Acad Sci USA} 98:404-409.


\bibitem{Bianconi:2008}
Bianconi G (2008) 
The entropy of randomized network ensembles. 
{ Europhys Lett} 81:28005.

\bibitem{Anand:2009} 
Anand K, Bianconi G (2009) 
Entropy measures for networks: Toward an information theory of complex topologies. 
{ Phys Rev E} 80:045102. 
 \bibitem{Latora_biased}
 G\'omez-Garden\~es J and Latora V (2008)
 Entropy rate of diffusion processes on complex networks.
 Phys. Rev. E 78:065102(R)
\bibitem{Eckmann:2004} 
Eckmann JP, Moses E, Sergi D (2004) 
Entropy of dialogues creates coherent structures in e-mail traffic. 
{ Proc Natl Acad Sci USA} 101:14333.

\bibitem{Song:2010}
Song C, Qu Z, Blumm N, Barab\'asi AL (2010) 
Limits of Predictability in Human Mobility. 
{ Science} 327:1018-1021.

\bibitem{Lambiotte}
Lambiotte R,   Blondel V D ,  de Kerchovea C, Huensa E,  Prieurc C,  Smoredac Z,  Van Dooren P(2008) 
Geographical dispersal of mobile communication networks,
Physica A, 387: 5317-5325.
\bibitem{Malmgren}
 Malmgren R D,  Stouffer D B,  Motter A E, and Amaral L AN (2008)
 A Poissonian explanation for heavy tails in e-mail communication
 Proceedings of National Academy of Science 47:18153-18158.
\end{thebibliography}
\end{document}